\documentclass[12pt,a4paper,reqno]{amsart}
\usepackage{amsmath,amssymb,amsthm, comment,graphicx}
\usepackage{tikz}
\usepackage{fancyhdr}
\usepackage{epsfig}
\usepackage{mathrsfs}
\usepackage{pifont}
\usepackage{amsfonts}
\usepackage{geometry}
\usepackage[compress,sort]{cite}
\usepackage[title]{appendix}
\usepackage[colorlinks,linkcolor=blue,anchorcolor=blue,citecolor=blue]{hyperref}
\allowdisplaybreaks
\newcommand{\RNum}[1]{\uppercase\expandafter{\romannumeral #1\relax}}
\numberwithin{equation}{section}
\numberwithin{figure}{section}

\usepackage{stmaryrd}
\usepackage{dsfont}
\usepackage{mathtools}
\usepackage{setspace}
\usepackage{tikz}
\allowdisplaybreaks[4]

\usepackage{bm}           
\usepackage{xcolor}
\usepackage{mathtools}
 
\usepackage{enumerate}
\usepackage[english]{babel}

\numberwithin{equation}{section}
\theoremstyle{plain}
\newtheorem{theorem}{Theorem}[section]
\newtheorem{lemma}[theorem]{Lemma}

\newcommand*\diff{\mathop{}\!\mathrm{d}}
\newcommand\numberthis{\addtocounter{equation}{1}\tag{\theequation}} 

\begin{document}

\title[A Principle of Maximum Entropy for the Navier-Stokes Equations]{A Principle of Maximum
Entropy\\ for the Navier-Stokes Equations}
\date{\today}

\author{Gui-Qiang G. Chen}
\address{Mathematical Institute, University of Oxford, Oxford,  OX2 6GG, UK}
\email{chengq@maths.ox.ac.uk}

\author{James Glimm}
\address{Stony Brook University, Stony Brook, NY 11794, and GlimmAnalytics LLC, USA}
\email{glimm@ams.sunysb.edu}

\author{Hamid Said}
\address{Department of Mathematics, College of Science, Kuwait University, Safat 13060, Kuwait}
\email{hamids@sci.kuniv.edu.kw}

\date{\today}

\begin{abstract}
A principle of maximum entropy is proposed in the context of viscous incompressible flow
in Eulerian coordinates. The relative entropy functional, defined over the space of $L^2$ divergence-free velocity fields,
is maximized relative to alternate measures supported over the energy--enstrophy surface.
Since thermodynamic equilibrium distributions are characterized by maximum entropy,
connections are drawn with stationary statistical solutions of the incompressible Navier-Stokes equations.
Special emphasis is on the correspondence with the final statistics described
by Kolmogorov's theory of fully developed turbulence.
\end{abstract}
\maketitle

\tableofcontents

\section{Introduction}\label{sec:intro}
We are concerned with the reformulation of the principle of maximum entropy,
as established in the context of statistical mechanics and information theory,
and its applications to mathematical fluid dynamics.
More specifically, the Boltzmann--Gibbs entropy 
under investigation ({\it cf}. \cite{Gavrilov17, Wherl}):
\begin{equation} \label{entropy-intro}
    S(f) = -\int f(u) \log f(u) \diff \eta(u)
\end{equation}
is maximized under appropriate constraints resulting from physical considerations\footnote{In case of a discrete system
in which probabilities $f_1, f_2, \cdots, f_k$
are assigned to a random variable $Y$ taking values $y_1, y_2, \cdots, y_k$, respectively,
then the entropy is expressed as $S = - \sum_i f_i \log f_i$.}.
From a mathematical standpoint, these considerations impose constraints on
the class of admissible velocity fields $u$ and their associated probability
distributions $f(u)$.

\smallskip
The principle of maximum entropy, introduced by E. Jaynes \cite{jaynes1957information} in 1957,
revolves around making the most desirable choice, for given prior data.
This choice aims to maximize the entropy functional \eqref{entropy-intro}.
In other words, as expressed in \cite{ihara1993information},
``[T]he [principle of maximum entropy] is the probability assignment that is consistent with the available
information but is maximally noncommittal with regard to missing information."
From the standpoint of statistical mechanics, it is seen as a method to determine
the correct probability distribution at the state of equilibrium,
depending heavily on the constraints imposed on the system, {\it i.e.},
our knowledge of prior or given information.
As the principle of maximum entropy provides a rule for assigning probabilities to data,
it proves particularly useful in ``filling gaps" between otherwise scattered data.
Therefore, maximum entropy modeling can be applied to a wide range of problems.
We refer to \cite{kapur1989maximum} for applications to several areas
including statistics, biology, and medicine.

\medskip
The physics underlying our problem is fundamentally rooted in the first and second laws of thermodynamics.
The former signifies the conservation of energy $E$,
while the latter imposes a constraint on admissible processes in terms of the production of entropy $S$.
At a micro-level, where each micro-state is equally probable,
entropy can be expressed in terms of the Boltzmann factor $k_B$:
$$
S = k_B \log (\Omega)
$$
where $\Omega$ is the number of states present at a given energy $E$.
Equation \eqref{entropy-intro} can be understood as a generalization of the above formula
for entropy $S$ when the micro-states of the system may not be equally distributed.

\smallskip
This formulation of the second law in terms of a single entropy can be rewritten in
a perhaps more conventional manner.
Introduce two entropies, $S_E$ for the entropy of the energetic degrees of freedom which is related to the viscous
and turbulent dissipation of energy, and $S_Z$ for the entropy of the vortical and enstropic degrees of freedom which
is related to the viscous and turbulent dissipation of enstrophy.
We assume that both the energetic and the enstrophic degrees of freedom are governed by a common temperature,
which can be fixed in the context of the second law of thermodynamics. Consequently, the second law takes the form:
$$
{\rm d}E = T {\rm d}S_E + T {\rm d}S_Z.
$$
The first of the terms on the right-hand side aligns with the familiar expression from
standard thermodynamics, a scenario that typically excludes physics models featuring an enstropic degree of freedom.
The inclusion of the second term is necessitated by the much richer physical model under consideration here.

\smallskip
In tandem with the modified or reformulated second law, we can express the Legendre transforms and the definition of entropy
as the logarithm of the volume of a constant energy surface, now accounting for two energy sources.
Consequently, we represent
$$
{\rm d}E = {\rm d}E_K + {\rm d}E_Z
$$
as a summation of the kinetic and enstrophic energies,
resulting in two terms on the left-hand side of the second law.
The kinetic energy $E_K$ aligns with the conventional description in thermodynamics,
while $E_Z$ is an additional energy source specific
to the effect that the entropy considered is for the simplest of the
fluid physics theories possible, namely incompressible with no mixtures.

\medskip
From a mathematical standpoint, the Boltzmann--Gibbs entropy (or its discrete counterpart) is a special case
of what is known as the \emph{relative entropy} first introduced by Kullback and Leibler 
in 1951 \cite{kull51} (consult \cite{kull-book} for more details).  
Consider a probability space $(\mathcal{H}, \mathcal{B(H)})$ and
two measures $\eta$ and $\mu$ defined on it, where $\mathcal{B(H)}$ denotes the space of Borel probability measures over
space $\mathcal{H}$.
Then the relative entropy $S(\mu | \eta)$ (of $\mu$ with respect to $\eta$)
is defined by
\begin{equation} \label{rel-ent-intro}
\mathcal{S}(\mu | \eta)
:=\begin{cases}
-\displaystyle\int_{\mathcal{H}} \dfrac{\diff \mu}{\diff \eta}
\log (\dfrac{\diff \mu}{\diff \eta}) \diff \eta(u)
& \text{if } \mu \ll \eta , \\
        - \infty &  \mathrm{otherwise}.
    \end{cases}
\end{equation}

In the case where measure $\mu$ is absolutely continuous with respect to $\eta$,
the Radon-Nikodym derivative $\frac{\diff \mu}{\diff \eta}$ is in $L^1({\rm d}\eta)$.
Once $\eta$ is chosen as the Lebesgue measure (when $\mathcal{H} = \mathbb R^n$),
then $\frac{\diff \mu}{\diff \eta}$ can identified with the probability distribution $f(u)$ in equation \eqref{entropy-intro}.
The relative entropy turns into a functional that is seen to be maximized over the space
of all probability measures, once we fix the \emph{reference} measure $\eta$.
Therefore, the maximum entropy problem consists of finding a probability measure $\mu$
on $(\mathcal{H}, \mathcal{B(H)})$, depending now on the choice of measure $\eta$,
such that the relative entropy is maximized under some given information.
In the probability space, $-\mathcal{S}(\mu | \eta)$ can be perceived as
a measure of ``{\it distance}'' between the maximum entropy measure $\mu$ and the reference measure $\eta$,
or any two measures for that matter.

\smallskip
In the context of fluid flow, a theory for entropy maximization was first established
in \cite{glimm2020maximum},
in which a well-defined probability measure acting as a linear functional
on the configuration space of the Euler and Navier-Stokes equations,
reduced to enforce particle interchange symmetries, was proved to maximize the Boltzmann--Gibbs
entropy relative to alternate solutions, restricted to the constant energy surface.
In this paper, we establish the existence of a probability measure on the configuration space
satisfying a maximum entropy admissibility condition relative
to alternate solutions of the Navier-Stokes equations,
yet with a restriction to the \emph{energy-enstrophy} surface.
The significance of this restriction lies in its connection to the pursuit of establishing
a rigorous mathematical foundation for the investigation of turbulent phenomena,
a link that we briefly outline in the following section.

\subsection{Motivation for entropy maximization}
The recent nonuniqueness result for the Navier-Stokes equations \cite{BucVic19}
suggest that weak solutions may fall short from describing the complex nature of fluid flow.
On this basis, a probabilistic description of the flow can serve as an alternate framework
for studying the Navier-Stoke and Euler equations, as opposed to classical weak solutions
and their Euler limits. In fact, any comprehensive exploration of turbulent flow phenomena 
must contend with some notion of \emph{averaging};
especially as it relates to formulating the Kolmogorov theory of fully developed turbulence, 
widely considered to be valid by the practitioners of the field.
We refer the reader to \cite{frisch1996turbulence} for a modern comprehensive account on this topic.

\smallskip
In an endeavor to establish a mathematical theory of turbulence,
Foias \cite{foiacs1972statistical} (see also \cite{foias2001navier}), inspired by earlier work of G. Prodi,
initiated a program to give, among other things, a rigorous formulation of the notion
of \emph{statistical solutions} for the Navier-Stokes equations and
eventually derive the predictions of Kolmogorov for the energy spectrum and the structure
functions \cite{kolmogorov1941dissipation,kolmogorov1991local}.
The Kolmogorov power laws are formulated in relation to
the \emph{average} (long time averages or ensemble averages) of certain physically
measurable quantities; hence, by constructing a family of time-dependent and spatially
homogeneous probability measures
$\mu_t$ (over an appropriate Hilbert space)
that satisfy the Navier-Stokes equations in some averaged sense\footnote{A closely related theory was 
introduced by M. Vishik and A. Furkisov  \cite{vishik2012mathematical},
where statistical solutions exist as probability measures on the set of solution trajectories of
the Navier-Stokes equations, {\it e.g.}, $C^{\rm w}_tL^2_x$. See also the work of Foias, Rosa, and Temam \cite{foias2013properties}
for a connection between these two notions of solutions.}.
Therefore, Foias, later with Manley, Roas, and Temam  \cite{foiacs1972statistical,foias2001navier},
was able to attach an unambiguous meaning to the notion of
an ensemble average\footnote{ In fact, Foias \emph{et al.} proved
the convergence of long-time averages $\lim_{t \to \infty} \frac{1}{t} \int_0^t \Phi(u(t)) \diff t$
for a suitable class of functions $\Phi$ and the weak solutions $u$ of
the Navier-Stokes equations. See \cite{foias2001navier} for the precise statement and more details.}.
In the context of this program, one main assumption embedded in the derivation of the Kolmogorov power 
laws is that the global energy \emph{identity} for the Navier-Stokes equations
(with normalized mass density and viscosity $\nu$):
\begin{equation} \label{glb-energy-id}
    \dfrac{1}{2} \|u(t)\|^2 + \nu \int_0^t \| \nabla u(s)\|^2 \diff s = \frac{1}{2}\|u_0\|^2
\end{equation}
is satisfied when integrated against a (spatially) homogeneous statistical solution $\mu_t$ for all $t > 0$.
Here $u_0$ is taken to be the (given) initial velocity of the fluid,
\emph{i.e.}, $u(x,0) \doteq u_0$, and $\| \cdot \|$ denotes the $L^2_x$-norm for simplicity.
These solutions, which also remain invariant under re-scaling, are called \emph{self-similar} homogeneous
statistical solutions to the Navier-Stokes equations.

\smallskip
Leray-Hopf solutions are known to satisfy the global energy \emph{inequality}
(obtained by replacing ``=" with ``$\leq$" in relation \eqref{glb-energy-id}).
Whether identity \eqref{glb-energy-id} holds in three space dimensions remains
one of the major open problems for weak solutions of the Navier-Stokes equations.
Incidentally, the same is true for statistical solutions: only an averaged version of
the global energy inequality is known to hold.

\smallskip
In other words, these claims infer that a probabilistic description consistent
with Kolmogorov's spectrum needs to be supported over the set
\begin{equation} \label{set}
 G_e(t) \doteq \Big\lbrace u\,: \,\dfrac{1}{2} \|u(t)\|^2 + \nu \int_0^t \| \nabla u(s)\|^2 \diff s = e(t) \Big\rbrace
\end{equation}
where the prescribed energy $e(t)$ is taken to be the energy profile for the initial velocity.
However, since the existence and uniqueness of such solutions remain in question, energy $e(t)$
can be prescribed \emph{a priori} to ensure that $G_e(t)$ is nonempty.
Throughout this paper, we assume that, for each $t\ge 0$, $G_e(t)$ is defined for any prescribed positive energy $e(t)$;
such a set $G_e(t)$ is called the \emph{energy-enstrophy surface.}

\bigskip
A similar challenge arises when constructing numerical solutions.
It is known that statistical solutions of the incompressible Euler equations
and multidimensional systems of conservation laws can be computed.
However, due to the lack of the well-posedness of these problems and the lack of convergence of their numerical approximations,
it is not known whether these solutions are entropy or entropy maximizing solutions\footnote{Incidentally,
as is mentioned in \cite{fjordholm2020statistical}, the lack of rigorous convergence of numerical schemes may be attributed
to the emergence of turbulent-like structures at finer scales as the mesh is refined.}.
See \cite{fjordholm2020statistical,lanthaler2021statistical} and references therein for more details.

\bigskip
Next, we demonstrate how the principle of maximum entropy provides a systematic approach for determining probability distributions
in the presence of constraints on the mean energy of the system.

\subsection{Principle of maximum entropy: classical formulation}
As indicated above, the principle of maximum entropy can be formulated in a number of different ways
depending on modeling considerations. The formulation in the context of equilibrium statistical
mechanics, which is presented here, most resembles the application of the principle
for the purposes of this work. Consider the entropy functional \eqref{entropy-intro}
together with the two constraints:
\begin{equation} \label{constr-intro}
  \int g(u) \diff \xi(u)=1, \qquad
  \int g(u) \,  \dfrac{m \,u^2}{2} \diff \xi(u)= \varepsilon,
\end{equation}
where $\dfrac{m \,u^2}{2}$ is the kinetic energy of a single particle as measured in a fixed frame,
$\varepsilon$ is the mean energy per particle which is \emph{a priori} known,
and measure $\xi$ is assumed to exist over the configuration space consisting of all velocity
fields\footnote{In the finite-dimensional setting, $\xi$ is taken to be the Lebesgue measure,
and the configuration space consists of all possible velocity components of a single
particle: $u = (u_1, u_2, \cdots, u_n) \in \mathbb{R}^n$.}.
The principle of maximum entropy, then, asserts that the physical entropy achieves its maximum, 
which we denote $g^*(u)$, for the constrained system. Hence we may apply
the methods of the calculus of variations to functional \eqref{entropy-intro}
subject to constraints \eqref{constr-intro} to conclude that
there exist two Lagrange multipliers $\alpha$ and $\beta$ satisfying
\begin{equation*} \label{EL1-intro}
 \int \dfrac{\delta L}{\delta g} (g(u); \alpha, \beta) \, \phi(u) \diff \xi(u) \bigg|_{g = g*}=0
\end{equation*}
for any test function $\phi(u)$, where
\begin{equation*} \label{Lag-intro}
    L(g(u); \alpha, \beta) = -g(u) \log g(u) + \alpha (1 - g(u) ) + \beta ( \varepsilon -g(u) \,  \dfrac{m \,u^2}{2}).
\end{equation*}
Therefore, we obtain
\begin{equation} \label{constr-intro-2}
-\int \phi(u) \Big( \big(1 + \log g^*(u)\big) + \alpha + \beta\dfrac{m \,u^2}{2} \Big) \diff \xi(u)=0.
\end{equation}
 If the integrand is assumed to be sufficiently regular, then
$$
\big(1 + \log g^*(u)\big) + \alpha + \beta\dfrac{m \,u^2}{2} = 0
$$
for all $u$, so that
\begin{equation} \label{EL-intro}
    g^*(u) = c_\alpha \exp(- \beta \frac{m \,u^2}{2}),
\end{equation}
where $c_\alpha = \exp(-(1+\alpha))$.
The two constraints \eqref{constr-intro} can be used to find
parameters $\alpha$ and $\beta$
in terms of the known mean energy $\varepsilon$.
One can even go a step further and determine the Lagrange multipliers
in terms of more fundamental thermodynamics quantities by employing
the first law of thermodynamics \cite{bais2007physics}.
In any case, we obtain the well-known Maxwell-Boltzmann distribution
describing the thermal equilibrium velocity distribution through the principle of maximum entropy.

\smallskip
In a similar fashion, the maximum entropy probability measure  $\mu_t$ (whose support is $G_e(t)$)
obtained for the problem of fluid flow should be associated with a universal law of decay
in the limit: $t \to \infty$.
This is indeed the case for self-similar homogeneous statistical solutions constructed:
the average kinetic energy scales like $\dfrac{M}{t}$ as $t \to \infty$,
where constant $M = M(\nu)$ is independent of any initial conditions ({\it cf}. \cite{foias2001navier}).

\smallskip
As will be illustrated in \S 2, a version of the energy constraint similar to \eqref{constr-intro}
will be presented, but not explicitly: it is included in terms of the support
of the maximum entropy probability measure, that is to say over set $G_e(t)$.
This choice is motivated by the role that the global (averaged) energy identity
for self-similar homogeneous statistical solutions plays in producing
the predictions of Kolmogorov's theory of turbulence.

\section{Entropy Maximization and the Physical Measure}

In this section, we establish the principle of maximum entropy in the context of incompressible flow.
We begin first by setting out the notation and formulating the problem.

\subsection{Principle of maximum entropy: formulation for the Navier-Stokes equations}

The fixed time particle configuration space is the Hilbert space ${\mathcal{H}}$ of $L^2$
divergence-free velocity fields defined on a cube $V \subset \mathbb{R}^3$
and satisfying periodic boundary conditions on the boundary of $V$.
For every $t \in [0,T]$ (fixed), define the Banach space $X_t$
to be the space $L^\infty_t L^2_x \cap L^2_t H^1_x$ endowed with the norm:
$$
\|u \|_t^2\doteq \dfrac{1}{2} \|u(t)\|^2 + \nu \int_0^t \| \nabla u(s)\|^2 {\rm d}s,
$$
where $\| \cdot \|$ denotes the $L^2_x$-norm.
Throughout this section, given a fixed $t \geq 0$,
we fix a scalar $e = e(t) \geq 0$ to define the energy-enstrophy surface:
\begin{equation} \label{2.1}
G_e(t)= \big\{u  \in X_t\,:\, \| u \|_t^2 = e(t) \big\}\qquad \mbox{for each $t\in [0, T]$},
\end{equation}
and make the identification
\begin{equation} \label{2.2}
G_e(0) = \big\{u_0 \in \mathcal{H}(V)\,:\, \|u\|_0 = \dfrac{1}{2}\|u_0\|^2= e(0) \big\}.
\end{equation}
It is clear that $G_e(t)$ and $G_e(0)$ are non-empty.
For each $t \geq 0$ and non-negative $e(t)$,
a given physical probability measure $\eta_{e,t}$ is supported over $G_e(t)$.
For measures $\eta_{e,t}$ and $\mu_{e,t}$ belonging the probability
space $\mathcal{M }:=(\mathcal{H, B(H)})$, we define
\begin{equation} \label{rel-ent-sec2}
\mathcal{S}(\mu_{e,t} | \eta_{e,t}) :=
\begin{cases}
- \displaystyle\int_{\mathcal{H}} \dfrac{\diff \mu_{e,t}}{\diff \eta_{e,t}}
\log (\dfrac{d \mu_{e,t}}{d \eta_{e,t}}) \diff \eta_{e,t}(u) & \text{if } \mu_{e,t}\ll \eta_{e,t}, \\
        - \infty &  \mathrm{otherwise},
\end{cases}
\end{equation}
which is known as the \emph{relative entropy} functional of measure $\mu_{e,t}$
with respect to the reference probability measure $\eta_{e,t}$. Since $\log(y)$ is a concave function
and $f_{e,t}(u) := \dfrac{\diff \mu_{e,t}(u)}{\diff \eta_{e,t}(u)} \in L^1(\eta_{e,t})$,
we employ Jensen's inequality to obtain
\begin{align*}
\mathcal{S}(\mu_{e,t} | \eta_{e,t})
= \mathcal{S}(f_{e,t})& = \int_{\mathcal{H}} \log( \frac{1}{f_{e,t}(u)} ) \diff \mu_{e,t}(u)
\le \log  (\int_\mathcal{H} \frac{1}{f_{e,t}(u)} \diff \mu_{e,t}(u)) \\
&\le \log ( \int_{\mathcal{H}} \diff \eta_{e,t}(u) )
=\log(\eta_{e,t}(G_e(t))) = 0.
\end{align*}
We postpone the discussion on the measure $\eta_{e,t}$ over the energy-enstrophy to the next section.

\smallskip
Our problem consists of maximizing the relative entropy functional \eqref{rel-ent-sec2}
subject to some appropriate constraints:
the solution obtained is a probability distribution
\begin{equation} \label{constr_1a}
\int    f_{e,t}(u)  \diff \eta_{e,t} (u) =\int   \diff \eta_{e,t} (u) = 1,
\end{equation}
and the support of the resulting measure must be supported on $G_e(t)$.
For the latter constraint, it will be shown that the fact that $\eta_{e,t}$
is supported on $G_e(t)$ is sufficient to ensure the second constraint is satisfied
by the maximum entropy measure. We hence require a solution to the following maximum problem:
\begin{equation} \label{MP}
    \sup  \big\lbrace \mathcal{S}(f_{e,t})\,:\,
    \mu_{e,t} \in \mathcal{M} \, \, \mathrm{satisfying} \, \, \eqref{constr_1a} \big\rbrace
\end{equation}
for each fixed $ t \in [0,T]$ and \textit{a priori} prescribed $e(t)$.

\subsection{Main theorem for the principle of maximum entropy}
The following basic lemma is key in establishing our main result -- Theorem \ref{Main_th} -- below.

\begin{lemma} \label{l1}
Fix $ t \in [0,T]$ and  $e(t) \geq 0$. Then the following statements hold{\rm :}
\begin{enumerate}
\item[\rm (i)] The relative entropy is non-positive{\rm :} $\mathcal{S}(f_{e,t}) \leq 0$
for all $\mu_{e,t} \in \mathcal{M}$, where $f_{e,t} = \dfrac{\diff \mu_{e,t}}{\diff \eta_{e,t}}$.

\item[\rm (ii)] The relative entropy is concave in $\mu_{e,t}${\rm :} for $\mu^1_{e,t}, \mu^2_{e,t} \in \mathcal{M}$
so that
$\mathcal{S} (\mu^1_{e,t} | \eta_{e,t})$ and
$\mathcal{S} (\mu^2_{e,t} | \eta_{e,t})$ are finite, then
$$
\mathcal{S} (\mu_{e,t} | \eta_{e,t})
\geq \alpha \mathcal{S} (\mu^1_{e,t} | \eta_{e,t})  + (1-\alpha) \mathcal{S}(\mu^2_{e,t} | \eta_{e,t}),
$$
where $\mu_{e,t} =  \alpha \mu_{e,t}^1 + (1-\alpha) \mu_{e,t}^2$ and $\alpha \in [0,1]$.

\item[\rm (iii)] The total variation of $\mu_{e,t}$, defined by
    $$
    |\mu_{e,t}|
    =\displaystyle\sup_{\Pi}\sum_{i=1}^m\int_{A_i} {\rm d}\mu_{e,t}(u),
    $$
    satisfies
    $$
    |\mu_{e,t} - \eta_{e,t}|^2 \leq -2
    \mathcal{S} (\mu_{e,t} | \eta_{e,t}),
    $$
    where the supremum is taken over all finite partitions $\Pi = \lbrace A_1, A_2, \cdots A_m \rbrace$ of $\mathcal{H}$.

\item[\rm (iv)] Functional $\mathcal{S}(f_{e,t})$ is upper-semicontinuous{\rm :} if $| \mu_{e,t}^n - \mu_{e,t} | \to 0$ as $n \to
\infty$, then
$$
\mathcal{S} (\mu_{e,t}| \eta_{e,t}) \geq \limsup_{n}\mathcal{S} (\mu^n_{e,t}| \eta_{e,t}).
$$
\end{enumerate}
\end{lemma}

\noindent
{\bf Proof}.
Statement (i) has been proven in \S 2.1. The proofs of (iii)--(iv) can be found in Ch.1 of \cite{ihara1993information}.

\smallskip
A slight modification of (ii) is also stated in  \cite{ihara1993information} but without explicit proof,
so we include the proof of (ii) for the sake of completeness.
We begin by invoking the inequality:
$$
\displaystyle\sum_{i=1}^m a_i \log(\dfrac{a_i}{b_i}) \geq a \log (\dfrac{a}{b}),
$$
where $a_i, b_i$ are any non-negative numbers for all $i = 1, 2, ..., m$, with $a = \sum_i a_i$ and $b = \sum_i b_i$.
The equality holds if and only if $\frac{a_1}{b_1} = \frac{a_2}{b_2} = \cdots = \frac{a_m}{b_m}$.

\smallskip
Fix $\alpha \in [0,1]$, and let  $\mu_{e,t} =  \alpha \mu_{e,t}^1 + (1-\alpha) \mu_{e,t}^2$. Consider
\begin{align*}
&\alpha \dfrac{d \mu_{e,t}^1}{d \eta_{e,t}} \log (\dfrac{\diff \mu_{e,t}^1}{\diff \eta_{e,t}})
+ (1-\alpha) \dfrac{\diff \mu_{e,t}^2}{\diff \eta_{e,t}} \log(\dfrac{d \mu_{e,t}^2}{d \eta_{e,t}}) \\
&= \alpha \dfrac{\diff \mu_{e,t}^1}{\diff \eta_{e,t}}
\log\left( \dfrac{ \alpha \dfrac{\diff \mu_{e,t}^1 }{\diff \eta_{e,t}}}{ \alpha \dfrac{\diff \eta_{e,t}}{\diff \eta_{e,t}}}\right)
+ (1-\alpha) \dfrac{\diff \mu_{e,t}^2}{\diff \eta_{e,t}}
\log\left(\dfrac{ (1-\alpha)
 \dfrac{\diff\mu_{e,t}^2}{\diff\eta_{e,t}}}{(1-\alpha) \dfrac{\diff\eta_{e,t}}{\diff\eta_{e,t}}} \right) \\
&\geq\left( \alpha \dfrac{\diff \mu_{e,t}^1}{\diff \eta_{e,t}} + (1-\alpha) \dfrac{\diff \mu_{e,t}^2}{\diff\eta_{e,t}}\right)
\log \left( \dfrac{ \alpha \dfrac{\diff\mu_{e,t}^1}{\diff\eta_{e,t}}
 + (1-\alpha) \dfrac{\diff \mu_{e,t}^2 }{\diff \eta_{e,t}}}{ \alpha \dfrac{\diff \eta_{e,t}}{\diff \eta_{e,t}}
  + (1-\alpha) \dfrac{\diff \eta_{e,t}}{\diff \eta_{e,t}}} \right)\\
    &= \dfrac{\diff\mu_{e,t}}{\diff \eta_{e,t}} \log ( \dfrac{\diff \mu_{e,t}}{\diff \eta_{e,t}}).
\end{align*}
Integrating with respect to $\eta_{e,t}$ gives the desired conclusion.

\smallskip
\smallskip
We now state and prove the main result of this paper.

\begin{theorem} \label{Main_th}
The physical measure $\eta_{e,t}$ is a solution to the maximization problem \eqref{MP}.
That is, the physical measure maximizes the entropy functional relative to
all the comparison probability measures $\mu_{e,t}$,
both restricted to the global energy-enstrophy surface $G_{e}(t)$.
\end{theorem}
\smallskip

\noindent
{\bf Proof}.  We divide the proof into two steps.

\medskip
1. \textit{Existence.} We follow the argument in Ch.3 of \cite{ihara1993information}. 
Fix $t \in [0,T]$ and a non-negative  function $e(t)$. Consider the problem
\begin{equation} \label{MIP}
     I_{\rm sup} = \sup_{\mu_{e,t} \in \mathcal{A}} \mathcal{S}(f_{e,t})\le 0,
\end{equation}
where 
$\mathcal{A} :=\lbrace \mu_{e,t} \in \mathcal{M} \, \, \mathrm{satisfying} \, \, \eqref{constr_1a}\rbrace$.
Let $\mu_{e,t}^n$ be a maximizing sequence for our problem, {\it i.e.}, $\mu^n_{e,t} \in \mathcal{A}$,
and $\mathcal{S}(f^n_{e,t}) \to I_{\rm sup}$ as $n \to \infty$. It can be directly checked that
\begin{align}
    \mathcal{S}(f^m_{e,t}) + \mathcal{S}(f^n_{e,t})
    &= \mathcal{S}(\mu_{e,t}^m | \eta_{e, t}) + \mathcal{S}(\mu_{e,t}^n | \eta_{e, t}) \nonumber\\[1mm]
    &= 2\mathcal{S}(\mu_{e,t}^{m, n} | \eta_{e, t}) + \mathcal{S}(\mu_{e,t}^m | \mu_{e,t}^{m, n}) + \mathcal{S}(\mu_{e,t}^n | \mu_{e,t}^{m, n}), \label{avg}
\end{align}
where $\mu_{e,t}^{m, n} = \dfrac{1}{2} (\mu^m_{e,t} + \mu^n_{e,t})$.
Because of the linear nature of the constraint, we see that
$\mu_{e,t}^{m, n} \in \mathcal{A}$.
The concavity of the relative entropy, Lemma \ref{l1}, gives
$$
2\mathcal{S}(\mu_{e,t}^{m, n} | \eta_{e, t})
\geq \mathcal{S}(\mu_{e,t}^{m} | \eta_{e, t}) + \mathcal{S}(\mu_{e,t}^{n} | \eta_{e, t})
$$
which, together with the non-positivity of $\mathcal{S} (\cdot)$, implies that
$\lim_{m,n} \mathcal{S}(\mu_{e,t}^{m,n} | \eta_{e,t}) = 2I_{\rm sup}$
so that
$\lim_{m,n} \mathcal{S}(\mu_{e,t}^m | \mu_{e,t}^{m, n})
= 0 = \lim_{m,n}  \mathcal{S}(\mu_{e,t}^n | \mu_{e,t}^{m, n})$.
By Lemma \ref{l1}(iii),
\begin{align}
|\mu_{e,t}^m - \mu_{e,t}^n|
&\leq | \mu_{e,t}^m - \mu_{e,t}^{m,n} | + | \mu_{e,t}^n - \mu_{e,t}^{m,n} | \nonumber\\[1mm]
&\leq  \sqrt{-2\mathcal{S}(\mu_{e,t}^m | \mu_{e,t}^{m, n})} + \sqrt{-2\mathcal{S}(\mu_{e,t}^n | \mu_{e,t}^{m, n})}
\to 0 \qquad\,\, \mbox{as $m, n \to \infty$}.
\label{cauchy-seq}
\end{align}
Then we conclude that there exists $\mu^*_{e,t} \in \mathcal{M}$ satisfying
\begin{equation}\label{3.9b}
\lim_n | \mu_{e,t}^n - \mu^*_{e,t}| = 0.
\end{equation}

Moreover, due again to the linearity of constraint \eqref{constr_1a}, it readily follows
that $\mu^*_{e,t} \in \mathcal{A}$.
Now, by the upper-semicontinuity of $\mathcal{S}$, Lemma \ref{l1}(iv), we conclude
$$
I_{\rm sup} = \mathcal{S}( \mu^*_{e,t} | \eta_{e,t}).
$$

\medskip
2. \textit{Euler-Lagrange equations}. Since the constraint is linear in $f$,
then, by the standard methods of the Calculus of Variations,
there exists a constant $\alpha_{t}$ (independent of $u$) satisfying the Euler-Lagrange equations:
\begin{align}
 - \int\Psi_t(u) \big(1 + \log(f^\ast_{e, t}(u))\big)  \, \diff \eta_{e,t}(u)
 - \alpha_{t} \int \Psi_t(u)  \diff \eta_{e,t}(u) = 0  \label{2.6b}
\end{align}
for any test function $\Psi_t(u)$.
Then we have
\begin{align}
- \big(1 + \log(f^{\ast}_{e, t}(u))\big) - \alpha_{t}
=0.  \label{2.7b}
\end{align}
for $\eta_{e,t}$-almost every $u \in G_e(t)$. Thus,
\begin{align*}
\log(f^{\ast}_{e, t}(u))=-(1 + \alpha_{t}).  \numberthis \label{2.8b}
\end{align*}
or equivalently
\begin{align} \label{2.9b}
f^{\ast}_{e, t} (u)
=  \exp \{- (1+\alpha_{t}) \}.
\end{align}
That is, $f^{\ast}_{e, t}$ is constant in $u$.
Since both $\mu_{e,t}^{\ast}$ and $\eta_{e,t}$ are probability measures, constraint \eqref{constr_1a} yields 
$\alpha_t \equiv -1$, implying that $ f^\ast_{e, t}\equiv 1$ or equivalently $ \mu_{e,t}^{\ast}=\eta_{e,t}$.
In other words, the physical measure $\eta_{e,t}$ achieves the maximum entropy
under constraint \eqref{constr_1a}.

\bigskip
{\bf Remark 2.1}.
The global energy-enstrophy surface $G_{e}(t)$ in \eqref{2.1} for Theorem \ref{Main_th} can be replaced
by the local energy-enstrophy surface: For any given $e(x,t)>0$,
\begin{align}\label{local-surface}
&\frac{1}{2}\int_V |u(x,t)|^2\psi(x,t)\diff x
 +\nu\int_0^t\int_V |\nabla u|^2\psi \diff x \diff \tau \nonumber\\
&-\frac{1}{2}\int_0^t \int_V \big\{|u|^2 ( \partial_t \psi + \nu \Delta \psi)
 +(|u|^2+2p)u\cdot \nabla \psi)\big\} \diff x \diff \tau
=\int_V e(x,t)\psi(x,t) \diff x
\end{align}
for any $\psi(x,\tau)\in C^\infty_{\rm c} (V\times [0,t])$.
Then Theorem \ref{Main_th} can state as follows: The physical measure maximizes the entropy production
rate relative to all the comparison probability
measures $\mu_{e,t}$, both restricted to the local entropy-enstrophy surface \eqref{local-surface}.

\bigskip
{\bf Remark 2.2}.
A connection can be drawn to the classical results highlighted in \S 1.2:
at $t=0$, we set $\eta_{e,0} = g(u)\, \xi$, where $g(u)$ is given by equation \eqref{EL-intro};
this implies that $\dfrac{\diff \mu^{*}_{e,0} }{ \diff \xi} = g(u) \in L^1({\rm d}\xi)$
in agreement with the classical principle of maximum entropy.
Furthermore, at $t = 0$, the energy condition in \eqref{2.2} gives the classical energy
constraint \eqref{constr-intro}$_2$, once we identify the mean energy
with $\int e(0) g(u) \diff \xi$.

\section{Connections and Remarks}
In view of the
previous work \cite{glimm2020maximum} on the principle of maximum entropy applied to the problem of inviscid fluid flow,
the current results outline a general approach for studying the case of viscous flow under a common framework.
As such, it is clear that the standard energy class associated with the Navier-Stokes equations provides the natural
setting for expressing the total energy  $\| u \|^2_t$ of the fluid at each time $t$,
and hence defining the energy-enstrophy surface $G_e(t)$.

\smallskip
Equally clear is that having a complete theory of entropy maximization for fluid flow
is tied with specifying selection criteria for the reference measure $\eta_{e,t}$.
Whether or not such criteria are model dependent is open to question.
For instance, in the finite-dimensional setting, a natural candidate for the reference measure is
the Lebesgue measure (see \cite{dunkel2007relative, glimm2020maximum}) as
it is the only measure (up to a multiplicative constant) that possesses the property of translation invariance;
that is, in the velocity space, it remains invariant under the Galilean transformation.
Yet another candidate for the reference measure is given by considering the statistics of fluid flows at a long time;
this is related to \emph{stationary} statistical solutions
of the Navier-Stokes equations and Kolmogorov's theory of turbulence ({\it cf}. \cite{foias2001navier}).

\smallskip
We first recall
the notion of the stationary statistical solutions of the Naiver-Stokes equations.
Assume that $\mu$ is a Borel probability measure over space $\mathcal{H}$ with finite enstrophy, that is,
$$
\int_{\mathcal{H}} \| \nabla u \|^2 \diff \mu (u) < \infty.
$$
Then the measure $\mu$ is said to be a stationary statistical solution of the Naiver-Stokes equations
if $\mu$ satisfies the following (stationary) Liouville-type equation:
\begin{equation} \label{stationary_NS1}
\int_{\mathcal{H}} \int_V Q(u) \, \Phi'(u) \diff x \diff \mu(u) = 0
\end{equation}
for all suitable test functions $\Phi$, and the energy inequality:
\begin{equation} \label{stationary_NS2}
    \int_{\mathcal{H}} \int_V \left( \nu | \nabla u |^2 - F \cdot u \right) \diff x \diff \mu (u) \leq 0,
\end{equation}
where $Q(u) = -\nu \Delta u + (u \cdot \nabla) u - F $.

\medskip
Now from the perspective of thermodynamics,
the flow relaxes, after sufficiently long enough time,
to its equilibrium state characterized by maximum entropy; so
we may assume that the statistics of this (final) state is encoded by the stationary solution  $ \mu \doteq \mu_{\infty}$.
On the other hand, one of the fundamental postulates of Kolmogorov's theory for fully developed isotropic turbulence
is the similarity hypothesis that the statistical behavior of the flow must be determined by only two parameters:
the (average) energy dissipation rate $\overline{\epsilon}_\nu(t)$ and viscosity $\nu$ (see also \cite{hopf1952statistical}).
Therefore, the average energy balance, in the absence of body forces,
for the case of non-zero energy dissipation rate ({\it i.e.}, the average of equation \eqref{glb-energy-id}) reads
\begin{equation} \label{energy-av}
    \dfrac{1}{2} \int_{\mathcal{H}} \|u\|^2 \diff \mu_t(u)
    + \underbrace{\nu \int_\mathcal{H} \int_0^t \| \nabla u(s)\|^2 \diff s \diff \mu_t(u)}_{\overline{\epsilon}_\nu (t)}
    = \frac{1}{2} \int_{\mathcal{H}} \|u\|^2 \diff \mu_0(u),
\end{equation}
where
the probability measure $\mu_t$ is such that equation \eqref{energy-av} 
is well-defined ({\it e.g.} a time-dependent statistical solution).
However, if we set $\mu_t = \mu_\infty$,
then the non-zero energy (finite) dissipation rate can be maintained only if the kinetic energy is infinite,
as seen by equation \eqref{stationary_NS2}. This nonphysical conclusion was pointed out
first by E. Hopf \cite{hopf1952statistical} (see also \cite{foias2001navier}).
The resolution of this paradox, according to Foias, Manley, Rosa,
and Temam \cite{foiacs1972statistical,foias2001navier,foias2013properties},
lies instead in introducing a \emph{family}
of probability measures  $\mu_t$ (parameterized by time $t \in [0, \infty)$)\footnote{More precisely,
for each homogeneous statistical solution of the Navier-Stokes equations $\mu_t$ (with viscosity $\nu$),
we define $\mu_t = \mu^{\nu, \overline{\epsilon}_\nu (t) }$ for some $\overline{\epsilon}_\nu (t) > 0$.},
a feature which the physical measure $\eta_{e,t}$ enjoys through the energy-enstrophy surface $G_e(t)$.
These solutions, called self-similar homogeneous statistical solutions, were originally postulated
for describing the decay of fully developed turbulence according to 
Kolmogorov \cite{kolmogorov1941dissipation,kolmogorov1991local},
though to date the question of their existence has not been resolved yet.

\smallskip
Meanwhile, our results suggest replacing the right-hand side of equation \eqref{energy-av} with $\overline {e}(t)$,
a quantity that can ostensibly be identified with $\int_{G_e(t)} e(t) \diff \eta_{e,t}$.
Such a choice adds a degree of freedom to the final statistics of the flow, now depending upon the initial prescribed energy $e(t)$.
In keeping with the above similarity hypothesis, however, we can in principle prescribe the total energy $e(t)$
at a high  Reynolds number in terms of parameters $(\overline{\epsilon}_\nu, \nu)$ and wavenumber $\kappa$,
removing the ambiguity associated with the term:  $\int_{G_e(t)} e(t) \diff \eta_{e,t}$.
Let $\mathcal{E}(\kappa,t)$ be the \emph{energy spectrum} (in the Fourier space with $ |\mbox{\boldmath$\kappa$}| = \kappa$)
associated with the average kinetic energy. Then the average total energy can be expressed in terms of the energy spectrum:
\begin{equation} \label{en-spec}
    \int_0^\infty \mathcal{E}(\kappa,t) \big( \dfrac{1}{2} +  \nu \kappa^2 \big) \diff \kappa.
\end{equation}

Equation \eqref{en-spec} provides a natural alternative for our (average) total energy $\overline{e}(t)$.
In fact, the amount of total energy contained between wavenumbers $\kappa_1$ and $\kappa_2$ at a given time $t$, \emph{i.e.},
$$
\int_{\kappa_1}^{\kappa_2} \mathcal{E}(\kappa,t) \big( \dfrac{1}{2} +  \nu \kappa^2 \big) \diff \kappa
$$
can now be identified with the change in the total energy $\overline{e}_{\kappa_2}(t) - \overline{e}_{\kappa_1}(t)  $.

\smallskip
As $\nu \to 0$, or equivalently at \emph{infinite} Reynolds number, the implications of Kolmogorov's theory give
\begin{equation} \label{kol-en}
\mathcal{E}(\kappa,t) \sim C(t) \, \overline{\epsilon}(t)^{\frac{2}{3}} \kappa^{-\frac{5}{3}}
\end{equation}
for large enough $|\mbox{\boldmath$\kappa$}| = \kappa$ but still belonging to the inertial range,
where  $C(t)$ is a positive dimensionless constant independent of viscosity $\nu$,
and $\lim_{\nu \to 0} \overline{\epsilon}_\nu (t) = \overline{\epsilon}(t)>0$.
In both cases, nevertheless, the explicit dependence on $e(t)$ can be removed
in the final statistics in favor of the energy dissipation rate and viscosity.

\medskip
In view of interpreting the total energy $e(t)$ in terms of the energy spectrum,
one can conceive of the maximum entropy probability measure,
that is the reference probability measure, as assigning a probability for finding the total energy carried
by a certain wavenumber $\kappa$ to be $\overline{e}_\kappa(t)$.
When the wavenumber lies in the inertial range and in the limit of infinite Reynolds number,
this probability can be calculated via the Kolmogorov energy spectrum \eqref{en-spec}.

\bigskip
\medskip
\textbf{Acknowledgments.}
Gui-Qiang G. Chen's research is partially supported
by the UK Engineering and Physical Sciences Research Council Awards
EP/L015811/1, EP/V008854/1, and EP/V051121/1.

\bigskip
\textbf{Declarations}

The authors have no competing interests to declare that are relevant to the content of this article. In addition, our manuscript has no associated data.

\end{document}